# Three research-based approaches to the design of physics activities

Amin Bayat Barooni,[1] Joshua Von Korff,[1] Brian D. Thoms,[1] Zeynep Topdemir,[1] Jacquelyn J. Chini[2]

[1]Department of Physics and Astronomy, Georgia State University, Atlanta, Georgia 30302, USA

[2]Department of Physics, University of Central Florida, Orlando, Florida 32816, USA

**Abstract**

A variety of activities are commonly used in college physics courses including lab, tutorials, and studio curricula. Instructors must choose among using research-based activities, designing their own activities or modifying existing activities. Instructors' choices depend on their own goals and the goals of activities from which they are choosing. To assist them in developing or modifying activities for their situation, we examine research-based activities to determine their goals and the features of the activities associated with these goals. Since most activities ask students to perform tasks to assist them in learning, sixty-six activities from eleven different research-based curricula were coded for student actions. The coding scheme containing 49 codes in ten categories was developed from a subset of activities, interviews with some of the activity designers, and recommendations from the American Association of Physics Teachers 2014 lab report. The results were examined using k-means cluster analysis revealing three design clusters. We label these clusters Thinking like a Scientist, Learning Concepts, and Building Models. These three clusters reflect diverse design goals. In the Thinking like a Scientist cluster, activities emphasize design of experiments by students, discussion, error analysis, reasonableness checking, supporting claims, and making assumptions or simplifications. The Learning Concepts cluster focuses on prediction of results and experimental observations. The Building Models cluster emphasizes discussion and answering physics or math questions that do not use collected data. This work connects common features appearing in physics activities with the goals and strategies of the designers. In this way it may provide instructors with a more straightforward way to create activities which achieve their desired outcomes.

**Introduction**

There is substantial demand for reforming physics curricula to include more interactive engagement activities and student learning improves by using these activities [1]. Activities come in many forms such as tutorials, labs, worksheets, and "ponderables"[2]. Sometimes these activities are used individually by instructors in lectures or laboratories and sometimes they are integrated into curricula such as SCALE-UP [2, 3], University Modeling Instruction [4] and Workshop Physics [5, 6].

andInstructors often turn to research-based activities as reliable methods that have been proven to help students learn. Activities are typically described as accounted research-based when results demonstrating their effectiveness have been peer-reviewed and published. For example, RealTime Physics (RTP) [7, 8], Socratic Dialog-Inducing (SDI) Labs [9-11], Cornell Thinking Critically in Physics Labs (CL) [12, 13], and Scientific Community Laboratories (SCL) [14] are examples of research-based activities.

However, there may sometimes exist a mismatch between the goals of the instructors and those embodied in the activities or instructional strategies. According to Henderson *et al*. [15] approximately one-third of instructors abandon the use of research-based instructional strategies (RBIS) after attempting one or more strategies. In other work, Henderson and colleagues [16] conclude that when faculty used RBIS "(i)n many cases they reinvented instruction that was missing important fundamental features of the intended instruction and/or conflicted with recommended practices." They report that instructors adapt and reinvent curricular materials due to "the personal nature of teaching and the unique instructional environments." In a project involving visits to ten U.S. universities teaching SCALE-UP physics courses, one of the authors (JV) observed that all but one institution used activities created by instructors rather than research-based activities. Many of these activities were observed to be research-inspired, meaning that the materials used principles developed through education research. Rather than using research-based activities as published, instructors chose to modify them or create their own activities to meet particular goals. The process of redesigning activities may be expensive and time-consuming, so developing principles that assist instructors in creating activities that meet particular pedagogical goals could be very useful. To design research-inspired materials, one needs to know useful design principles, and the efficacy of the materials will depend entirely on the efficacy and accuracy of those principles. The intent of this publication is to help instructors design research-inspired activities by providing them with those principles [17, 18].

Meltzer and Thornton reviewed many active-learning instructional methods in physics and discussed their effectiveness for student learning [19]. They also identified some common characteristics among research-based active-learning instruction in physics such as student ideas are elicited and addressed, students express their reasoning explicitly, qualitative reasoning and conceptual thinking are emphasized, problems are posed in a wide variety of contexts and representations, instruction frequently incorporates use of actual physical systems in problem solving, and instruction emphasizes linking of concepts into well-organized hierarchical structures. The present work identifies common characteristics of research-based physics activities but also examines the connection of these characteristics to particular design goals. Design goals for activities reported in the literature include learning concepts [4, 6, 8, 9, 20, 21], thinking like physicist or scientists [22, 23], understanding measurement and uncertainty [12, 14], designing experiments [12, 14, 20], constructing models [4, 24], and improving students' beliefs about the nature of experimental physics [25].

Our long-term research goal is to find a practical way to help instructors who want to use the features of different research-based activities to design their own activities. The present work takes a primary step toward this goal by grouping research-based activities according to their design features. This work seeks to answer the following research questions:

- How can we cluster the research-based activities based on their design features?
- How are these clusters related to the goals of the designers?

Preliminary results from this project have been reported in which we investigated several uses of representations in evidence-based and non-evidence-based physics activities [26], the role of revisiting as an essential and common technique in tutorials [18], and analyzing several design philosophies revealed through interviews with designers of research-based activities [17].

## Methods

**Sources**

We used three primary sources to develop the coding scheme in this work: 1) interviews with the designers of research-based activities, 2) 2014 AAPT lab report, and 3) published research-based activities. The coding system was then used to characterize research-based activities from eleven research-based curricula. To arrive at reliable conclusions from these three sources, we used a triangulation approach in our qualitative research to analyze the data from different perspectives [27].

1) **Interviews**

We performed one-hour, semi-structured interviews with 15 designers of some of the most frequently used research-based activities including Investigative Science Learning Environment (ISLE), University Modeling Instruction (UMI), RealTime Physics (RTP), Workshop Physics (WP), Tutorials in Introductory Physics (TIP), and Open Source Tutorials (OST). The main focus of the interviews was to ask about the principles or techniques they used to design activities, how these design principles help students to achieve the goal of the activities, and what the similarities and differences are between their activities and other commonly-used activities. Our aim was to learn more about the goals and purposes of designers than is revealed in published articles. Participants expect that we will not reveal their names, but that they may be identifiable based on there being only one or a few authors for each curriculum. The interviews were video recorded, transcribed by one researcher, and the transcriptions were checked by another researcher for accuracy.

2) **AAPT lab report 2014**

In 2014 American Association of Physics Teachers (AAPT) published *Recommendations for the Undergraduate Physics Laboratory Curriculum* [28] suggesting specific learning goals for introductory physics laboratories such as constructing knowledge, modeling, designing experiments, developing technical and practical laboratory skills, analyzing and visualizing data, and communicating physics.

3) **Research-based Activities**

In this work we evaluated 66 introductory college level laboratory and classroom activities from 11 research-based curricula. Most of these were chosen based on their effectiveness as reported by Von Korff *et al*. [29] who evaluated the results of Force Concept Inventory and the Force and Motion Conceptual Evaluation for different interactive engagement teaching techniques published between 1995 and 2014. Activities from two research-based curricula were included because they were studied in earlier work by Thacker *et al*. [21]. In that study FCI gains from "physics education research-informed materials" were compared to traditional activities in a large university. Two additional resources were included because they are widely-used research-based activities recommended on the PhysPort website [30]. Some of the activities were free to download, and others required that we receive permission from the authors or designers. The sets of activities investigated are shown in Table I. Among the activities studied here, ISLE is not only a set of activities, but also a design philosophy; so there may be many labs that are compatible with ISLE. We used a particular set of labs created by the authors of the ISLE design philosophy.

**Analysis**

Figure 1 shows a diagram representing our coding process. Since activities by their nature ask the students to perform tasks which are intended to help them learn, we coded the designer interviews for expected student actions in order to understand the goals of the activities. We used the constant comparative method to develop the coding scheme and we met regularly to discuss and debate the codes and their arrangement into categories (axial coding) [31]. We used this scheme to code 25 different activities (referred to as Set 1), mainly from the same research-based curricula as the interviewees. At each step the coding scheme was validated for consistency using Cohen's Kappa to evaluate inter-rater reliability (IRR) [32]. For the IRR process, we count codes one time if we observed them in the minimum possible unit for coding. For example, if this minimum possible unit is a paragraph and we observe a code three times, we count it only once. The reason behind this policy is that consecutive features of the same type within a paragraph were generally closely related. After finishing the IRR process, the researchers discussed the results, then eliminated or combined codes and rewrote code definitions as needed. Memos helped us to write our thought process and refer to them in next steps [33]. Preliminary results from this work were published in 2016 [18]. The code list from this first stage was used to analyze the introductory level recommendations in the AAPT lab report 2014 and 40 additional activities chosen from all eleven research-based curricula (Set 2) and to reanalyze the designer interviews. Additional results were reported by the authors at this stage of the investigation [17]. Again, the researchers discussed and revised the code list and categories informed by a determination of IRR. The code list and categories were then evaluated by two independent physics education researchers to provide feedback. Small changes in the categories and their constituents were made in response to this feedback.

The final scheme consisted of 49 codes in ten categories as shown in Table II. One activity was randomly chosen from each research-based curricula for coding and evaluation of IRR. This evaluation yielded an average Cohen's kappa of 0.78 per activity. According to Everitt [34] this value of Cohen's Kappa is "satisfactory or solid agreements", and according to Fleiss *et al.* [35] it shows "excellent agreement". The minimum and maximum values of Cohen's kappa we obtained were 0.653 and 0.925, respectively.

After achieving good IRR results, one of the researchers randomly selected five additional research-based activities from each research-based curricula for coding. The coding results from six activities from each of the eleven research-based curricula were then used for the cluster analysis.

**k-means cluster**

The frequency at which codes appeared varied greatly among the activities and research-based curricula we analyzed. We applied a k-means cluster analysis [36] to group the activities according to the pattern of codes. According to Formann [37], applying a k-means cluster analysis to data with *m* features requires a minimum of $2^m$ data instances. Since we analyzed a total of 66 research-based activities with our final coding scheme, we were limited to choosing six features for the cluster analysis. Of the 49 codes in ten categories, we identified one code (Non-observation Questions) and five categories (Observations, Prediction, Spoken Representation, Design, Qualifications) as the best features for the cluster analysis (highlighted in Table II). Five categories appeared in our analysis with high frequency and also showed significant variations among the activities. For example, the Written Representation category had high frequency but showed little variation among the activities since this is an extremely common expectation for student action. In contrast, the Design category appeared with high frequency in some activities and rarely in many others. The only single code selected as a feature for the k-means cluster analysis, Non-observation Questions, was coded when students were asked a physics or math question that did not use data from a previous measurement or observation and was not a prediction of future observations.

The k-means cluster analysis was performed in Python (Jupyter Notebook version 5.7.8) using the KMeans function in the sklearn.cluster package. Each of the 66 activities were points in the analysis and the Euclidean distance was used to measure similarities among points [38]. We define $v_{ij}$ to be the value of the frequency for the $i^{th}$ feature in the $j^{th}$ activity. For instance, if an activity had three of code X and one of code Y and no other codes, their frequencies would be 0.75 and 0.25, respectively. Finally, because some student actions are more prevalent than others, we normalize each feature's frequency over all activities using z-scores to determine the final values of $v_{ij}$. These frequencies locate each of the 66 activities in a six-dimensional space. The goal of k-means cluster analysis is to locate the N cluster centers that minimize how far activities are from their cluster center. The center of each cluster C is defined as $v_{iC} = \frac{1}{N_C} \sum_{j \in C} v_{ij}$ where the sum is taken over all activities in the cluster, and $N_C$ is the number of activities in the cluster. Each activity is taken to be a part of the cluster whose center is closest to that activity using the Euclidean distance, $D_j = \sqrt{\sum_{i=1}^{6}(v_{ij} - v_{iC})^2}$, between them in a six-dimensional space. The clustering process begins by selecting *N* initial centers at random and determining the activities that are nearest those centers. After each activity is assigned to a cluster, a new center for each cluster is computed. Since this may cause some activities to now be closer to the center of a different cluster, the distances $D_j$ are recalculated and each activity is again assigned to a cluster and a new cluster center computed. This process repeats until there is no longer any change in the cluster assignments.

However, this result could be a local optimum solution rather than a global optimum. Therefore, the process is repeated 1000 times with new randomly chosen initial centers. For each repetition the quality of the clustering is evaluated by calculating the inertia of each clustering solution, $I = \sum_j D_j^2$. The cluster arrangement with the lowest inertia is taken as the final solution for each value of *N*. Since the best choice for the number of clusters is not known, the "elbow method" [39] was used. To apply the elbow method, we computed the inertia $I_N$ for each number of *N* clusters, with *N* between 1 and 5. The inertia is a measure of the quality of fit of the clustering with a larger inertia meaning a worse fit. As such, IN should decrease when new clusters are added and N is increased. As shown in Fig. 2, the inertia gradually decreased as the number of clusters increased from one to five as expected. The elbow method determines the relative improvement of the fit by the addition of the $N^{th}$ cluster by maximizing $(I_{N-1} - I_N) - (I_N - I_{N+1})$. When this value is large, it means that *N* clusters produce a much bigger improvement over *N-1* clusters than *N+1* produces over *N* clusters, suggesting *N* clusters as the optimal choice. Figure 3 shows the application of the elbow method to our data and reveals that the largest improvement is achieved when adding the third cluster.

## Results

To determine the design goals exhibited by each of the three clusters, we examined each cluster for patterns in the frequency of coding features. We evaluated those patterns by comparing them with design goals expressed in the literature and interviews. Figure 4 shows the average z-score of normalized frequencies for each feature for the three clusters. This analysis led us to name the three clusters Thinking like a Scientist, Learning Concepts, and Building Models, as described in more detail below. While we placed each activity in one of the clusters, all activities will do each of those to some extent, of course. But the clusters reveal that individual activities embody one goal more than the others and therefore lean more heavily on particular student actions to accomplish the goal. Statements from the activity designers, in literature or interviews, also reveal these goals.

### Thinking like a Scientist

Compared with the other clusters, the most frequent features observed were Design, Produce Spoken Representation, and Qualifications (as described in Table I). These features show an emphasis on students performing scientific practices such as experiment design, reaching decisions by collaboration, and examining results and processes for accuracy (such as error analysis, checking assumptions and simplifications, and evaluating reasonableness). Activities in this cluster include all six CL, five SCL, and three ISLE labs (Table I).

In recent years there has been an increasing focus on explicitly teaching scientific thinking in physics courses. Holmes *et al*. [13] argue that students need to develop quantitative critical thinking and that developing this requires "repeated practice in making decisions based on data, with feedback on those decisions." Etkina and Planinšic [23] explain that the "ability to think like a scientist while solving complex problems is … vital ." They state that students need to be able to "formulate a problem; collect and analyse data; … identify patterns., … test ideas; … evaluate assumptions and solutions; ... distinguish evidence from inference; … argue scientifically."

According to Lippmann, the main goal of SCL is to teach "skills and techniques for creating, transforming, and evaluating scientific knowledge" [14] and that students "understand the concepts underlying uncertainty in an experiment (called measurement concepts) and be able to use that knowledge to design an experiment and interpret their data." Holmes *et al*. state that one of the goals of the CL is "thinking like a physicist" where students gain scientific skills to apply data to "evaluate models, explanations, and methods" [22]. Etkina *et al*. report that designing of an experiment by students is one of the critical components of the ISLE philosophy [20]. They also state that students in the ISLE classroom engage in the process's scientists use to achieve knowledge by collaborating in groups and sharing ideas. According to our interviews the designers of ISLE regard Thinking like a Scientist as an important goal of ISLE [17]. Holmes and Wieman previously pointed out that ISLE and CL both focus on making decisions in the experimentation process asking students to evaluate their outcomes [22].

The most distinct feature of the Thinking Like a Scientist cluster compared with the other two clusters is the prevalence of the Design and Qualifications features. These two features emphasize student decision-making in the creation, modification, and evaluation of experimental methods. The interviews and publications from designers make clear that their goal is for students to develop the process skills used in a scientific approach. Although Spoken Representations is prominent in both the Thinking Like a Scientist and Building Models clusters, the reasons appear to differ. Designers of activities in the Thinking Like a Scientist cluster talk about scientific arguments and explanations and emphasize students' critical evaluation of their own and each other's ideas.

**Learning Concepts**

Compared with the other clusters, the most frequent features observed were Prediction and Observation (Table I). The Learning Concepts cluster is the largest of the three clusters and includes all six analyzed activities from RTP, WP, Physics Department, Texas Tech University (TTU), and The University of Illinois labs (UI). This cluster also includes five of the SDI Labs, three activities each from ISLE and TIP, two from University Modeling Instruction, and one from each of SCL and OST (Table I).

Sokoloff *et al*. report that two purposes of RTP are to support students to "acquire an understanding of a set of related physics concepts" and to "master topics covered in lectures and readings using a combination of conceptual activities and quantitative experiments" [8]. According to our interview with one of the RTP designers, they achieve this goal by using a learning cycle of prediction, observation, and comparison. Interviews also revealed that this learning cycle was used by WP. According to Laws, learning concepts is one of the main goals of WP to help students to succeed in physics, engineering, and sciences [40]. According to Thacker *et al*., teaching concepts is an important factor in PER-informed labs such as those used at TTU [21]. They report that these labs are designed to "to address common student difficulties and conceptions by posing appropriate questions to elicit, confront, and resolve the difficulties." They also report that these labs let students "make observations that might challenge or contradict their present conceptual understanding and allow them to reshape their conceptual understanding through thought and discussion." Thacker *et al*. also state that "UI were designed as part of the reform of their introductory courses and were designed as an adaptation of the approach of Real Time Physics, designed to address common misconceptions through active engagement of the students in the learning process." According to Hake, the primary goal of SDI is "to promote students' mental construction of concepts" [11].

Learning concepts is one of the main goals of TIP according to Kryjevskaia *et al.* who report that the "overarching goal of the tutorials is to promote functional understanding of concepts that are challenging for many students even after traditional instruction" [41]. Interviews with the designers of TIP revealed more details of the design approach and their use of Elicit-Confront-Resolve as a strategy for learning concepts. A designer explained the role of predictions and observations in student learning as a way for students to see a "confrontation between the way they were thinking and the prediction that would logically from that model, and yet the experiment – nature, disagrees."

Etkina and coworkers give constructing physics concepts as one of the main goals of the ISLE [42] and explain the role of prediction and experimentation in their learning cycle [20, 42, 43]. They explain that the ISLE process starts with students observing an initial experiment, then after constructing explanations they test their model with predictions and further experiments. Students may then modify and/or abandon their explanations and perform additional experiments. In interviews the designers of ISLE explain that one "can think of observational experiments as concept building experiments, ... testing experiments are concept testing experiments, you need to test it, and application experiments are multiple concepts that you have tested."

The goals expressed by the designers of activities in the Learning Concepts cluster appear consistent with the key tenets of conceptual change theory. The most frequent features show an emphasis toward students expressing their conceptual understanding and collecting data to test their ideas. González-Espada *et al*. report that prediction and comparing the result of prediction with observation helps students change their conceptual understanding [44]. According to Chi, using prediction and testing allows students to successfully modify their mental model [45]. Khourey-Bowers states that using predictions and hypothesis generation is one of ten strategies for conceptual change instruction which can "awaken curiosity and inspire questioning" [46]. Hesse claims that an important step in conceptual change is challenging conceptions in which students predict according to non-scientific concepts followed by a demonstration event and explanation of the correct answer [47]. So, it's not surprising that the key features of this cluster, prediction and observation, are those which bring out students' pre-conceptions, require comparison with results of experiments, and confirm or refute their understanding. While the goal of these activities may be the construction of new mental models, the approach differs from the cluster we have labeled Building Models in that it relies on physical experimentation and observation. This is consistent with the idea of creating dissatisfaction through a "discrepant event" in conceptual change theory [44].

**Building Models**

Compared with the other clusters, the most frequent features observed in this cluster were Spoken Representation and Non-observation Questions (Table I). Non-observation Questions are physics or math questions that do not use data from a previous measurement or observation and are not prediction questions. They tend to engage students in problem-solving, refining their intuitions, using and interpreting representations, and model building, e.g. students try to prove a formula or make a hypothesis. Activities in this cluster include five OST, four UMI, three of the TIP, and one SDI lab (Table I). According to David Hestenes [48] "models in physics are mathematical models, which is to say that physical properties are represented by quantitative variables in the models."

Lising *et al*. [49] report that a goal of OST is student model building. According to interviews, OST activities are designed with explicit attention to the metacognitive and epistemological aspects of student learning. One aspect of this is students' revision of incorrect answers by a process of refining intuition and reconciliation. The designers explained that some OST activities use a lab without predictions since the goal of these activities is not doing experiments but instead to help students find a pattern and use more mathematical reasoning. One designer described the OST activities as having a "sense-making" feature. One of the designers of OST explained in an interview that students' spoken representations were important in the model building process. A designer states that as part of this epistemological process students are required to talk to a TA or instructor at particular points in the activity because they "wanted to provide opportunities for students to think [about] their thinking [and] instructors to engage students in those conversations and make that explicitly a part of the exercise."

According to the interview with designers of UMI, modeling is the process of building, testing, validating, and revising models. The purpose of labs is "not to confirm something that we have introduced theoretically, it is instead to introduce a new phenomenon." So, they state that "labs often are very conceptual and oriented around introducing something and bringing about a change in the modeling cycle." Brewe [4] reports that problem solving in modeling instruction differs from traditional problem solving since it is about "the application and adaptation of models." According to interviews with designers of UMI, "models are built up of representations" including spoken representations. In UMI activities, students perform "white board discussions" where students share their individual models and modify them.

TIP activities coded in this work were divided between the Learning Concepts and Building Models clusters. While some TIP activities ask for predictions followed by small experiments, according to interviews with designers some activities do not require an experiment but instead ask questions aimed only at having students construct a model. One of the designers describes their idea of model building as "breaking something up into constituent pieces and sometimes it's sort of representing sort of a complex thing." A designer explained about the goal of building models in TIP as "they (students) want to have a sort of procedure that they can say, how can I build a prediction based on think(ing) of these wave as if they are like there's this fictional pulse or, how can I predict what an extended light source, what kind of image an extended light source is going to produce based on thinking of it as many tiny sources. Or, how can I think of a circuit if I think of it as something flowing and there's pathways and barriers to that flow." According to interviews, spoken representation is one of the design principles of TIP where students are required to have discussions about their ideas in groups and at points to check with the instructor to make sure they resolve their inconsistencies. One designer explained that TIP questions are meant to be difficult for a student to answer alone which encourages students to participate in the group discussion. A key feature of the Building Models cluster is the use of non-observations Questions which ask students to rely on mathematical or physical reasoning rather than observations. While the ultimate goal of achieving a new mental model may be similar to activities in the Learning Concepts cluster, the methods often differ. In some cases, it may be that the concepts involved do not lend themselves to direct observation but are more accessible to a mathematical approach. In other cases, it may be that the underlying goal is for students to develop the sense-making, metacognitive, and epistemological skills they need to evaluate their framework of ideas. The designers emphasize conceptually complex problems which may require more of these skills. This process of developing sense-making, metacognitive, and epistemological skills and applying them to more complex situations appears to be the main motivation for the prevalence of the producing spoken representations feature in the Building Models cluster.

## Study limitations

One limitation of this study lies in the issue of learning cycles. Activity designers in some cases order their activities to build skills over a sequence. In this study the unit of evaluation for the cluster analysis was individual activities which were chosen randomly from the available materials from each research-based curriculum. This investigation was not designed to capture skill-building on longer scales. For example, designers of UMI stated in interviews that the learning cycle consists of a unit of instruction rather than a single activity. They state that "modeling is definitely slower, and you have to make a lot of choices of like, what content coverage versus, like, breadth versus depth."

## Conclusions

To assist instructors who want to develop or modify their activities associated with their goals by using the features of different research-based activities, we coded student actions in sixty-six activities from eleven research-based curricula and analyzed code frequencies using k-means cluster analysis. The result of this analysis was three clusters.

1) **Thinking like a Scientist** cluster's most important features are designing experiments by students, spoken representation, error analysis, reasonableness of student's answers, assumptions, simplifications and limitations. These activities focus attention on students performing scientific practices. These features are supported by design principles that focus mainly on designing experiments and evaluating the results.

2) **Learning Concepts** cluster mainly concentrates on observation and prediction. Activities in this cluster emphasize conceptual understanding of students and collecting data from experiments to test their hypothesis. This cluster uses essential points of conceptual change theory.

3) **Building Models** cluster focuses on tools for helping students to solve conceptually complex problems. It's two most prominent features are spoken representation and non-observation questions where students address physics or math questions without using collected data or observations. Activities in this cluster tend to engage students in problem-solving, refining their intuitions, using and interpreting representations, and model building, e.g. students try to prove a formula or make a hypothesis.

In this work we have identified connections between features that appear in physics activities and the goals of the designers. Making explicit the connections between the design goals and the activity features may provide instructors with a better way to select from among published activities and also lay out a clearer path to create new activities to address the learning goals they have for their students. In this way instructors may be able to create physics activities with a more consistent design philosophy.

## Acknowledgement

The authors gratefully acknowledge financial support from the National Science Foundation Award No. 1347510. We thank the designers who gave their valuable time to participate in this study. We also thank Jon Gaffney for his helpful suggestions and Monica Cook, Hannah Pamplin, Myat Pho, Ibraheem Robins, Brian Ferguson, and Kyle Simmons for their collaboration during the research.

**Table I: List of 11 research-based curricula whose activities were investigated and numbers of activities in each cluster from each research-based curricula.**

| List of research-based curricula | Thinking like a Scientist | Building Models | Learning Concepts |
|---|---|---|---|
| Cornell Labs (CL) Version 2018, received in person [12] | 6 | 0 | 0 |
| Open Source Tutorials (OST) [50] | 0 | 5 | 1 |
| University Modeling Instruction (UMI) [51] | 0 | 4 | 2 |
| Workshop Physics (WP) [52] | 0 | 0 | 6 |
| Socratic Dialog Inducing Laboratories (SDI) [10] | 0 | 1 | 5 |
| Tutorials in Introductory Physics (TIP) [53] | 0 | 3 | 3 |
| Physics Department, Texas Tech University (TTU), received in person [21] | 0 | 0 | 6 |
| The University of Illinois (UI), received in person [21] | 0 | 0 | 6 |
| RealTime Physics (RTP) [54] | 0 | 0 | 6 |
| Investigative Science Learning Environment (ISLE) [55] | 3 | 0 | 3 |
| Scientific Community Laboratories (SCL) [56] | 5 | 0 | 1 |

**Table II: List of the all the codes and categories with the six features used in the k-means cluster analysis highlighted.**

| Category | Code | Definition |
|---|---|---|
| Observation: The process of collecting data by observation and using it | Observe Data from Equipment | Refers to a physics or math question or instruction about observation and recording of data from equipment by students. |
| | Observe Data from a Simulation | Refers to a physics or math question or instruction about observation and recording of data from simulation by students. |
| | Use Observed Data from Equipment | Refers to a physics or math question about data previously observed and recorded from equipment by students. |
| | Use Observed Data from Simulation | Refers to a physics or math question about data previously observed and recorded from simulation by students. |
| | Extract Data from Video | Asks students to extract data from videos during the observation. |

| Prediction: The process of prediction an experiment or making hypothesis by students and comparing the result of the experiment with the prediction | Make Prediction | Asks students to make a prediction, which means that (1) an experiment will be done in the future and (2) the students are asked to figure out the result before experimenting. |
|---|---|---|
| | Check Prediction | Asks students to decide if a prediction was consistent with their observation. |
| Spoken Representation: Communication among students, instructor, and class | Check with TA or Instructor | Asks students to talk to the instructor about some work they have been doing. |
| | Group discussion | Asks students to talk to their group. |
| | Class discussion | Asks students to talk to the whole classroom. |
| | Symposium | Asks students to visit or talk to other groups. |
| | Show Whiteboard | Asks students to show their whiteboard to an instructor, another group, or the whole class. |
| | Think/Pair/Share | Refers to thinking individually, comparing answers with other group members and resolving any conflicts. |
| Design: The process of design, improvement, making hypothesis for an experiment or math procedure by students | Procedure Design | Designing procedure, could include describing an experimental procedure invented by the students, explaining how the students invented an experimental process, or explaining what decisions the students had to make to invent the experimental procedure. |
| | Improve design | Asks students to improve their previous designs. |
| | Choose Question to Investigate | Asks students to choose an open-ended inquiry question. |
| | Designing Math Procedure | Asks students to design/state/invent/improve a mathematical or quantitative procedure they will use before they use it. |
| | Making Hypothesis | Asks students to make a hypothesis that they have devised. |
| Qualifications: Asking students for their assumptions, simplifications, limits, error analysis and reasonableness of their answers | Assumptions, Simplifications, Limits | Asks students about their assumptions, simplifications, and limitations with their model or way of understanding a physical situation |
| | Error and Uncertainty | Asks that the students either give a qualitative discussion of error or estimate or quantify the uncertainty or error. |
| | Reasonableness | Asks about the reasonableness of results or answers. Also, includes questions about the nature of reasonableness, or what it means for something to be reasonable |

| | Most Important Concept | Asks students what the most important concepts are |
|---|---|---|
| | Goal or Purpose | Asks about what students will have or be able to do or what question the students will answer by the end of the lab. |
| | Non-observation Question | Students answer physics or math questions that do not use data from a previous measurement or observation and are not a prediction question. |
| | Instructor Guide | Tell the instructor what to do (as opposed to telling students what to do). The activity might tell the instructor to lecture in a certain way, to help students' groups, or to give a demonstration |
| | Real-world Example | Asks students for a real-world example. |
| Miscellaneous: Any code not in another category | Computer Data Analysis | Asks students to use a computer to analyze existing data numerically. The computer processes and displays numbers or equations such as errors, means, or the parameters of a curve fit. |
| | Grading Rubric | Gives a grading rubric to students showing how students' work will be assessed. |
| | Generalization | Asks students to Identify trends or reason by induction to produce generalizations. |
| | Calibration | Students are instructed to calibrate equipment, such as a scale. |
| | Ethics | The instruction mentions ethical considerations, such as plagiarism. |
| | Notebook | Students should write something in a lab notebook, report, or other documents that are separate from the activities' questions. That means students have to organize their responses themselves. |
| Epistemology: Refers to epistemological questions | Epistemological Question | The activity asks general questions about how to think, how to learn, how to proceed with certain kinds of physics problems, or how to go about doing physics. |
| Written Representation: Different kind of written representations | Written Word | Asks students to write their idea or explain something. |
| | Math | Asks students to write variables, numbers, equations, and units. |
| | Student Chosen Representation | Asks students to produce a model, choose one or more representations, or to use multiple representations, but the |
| | | specific representations are not named. The students are not told what to do with representations. |
| | Diagram | Asks students to draw a diagram. |
| | Graph | Asks students to draw graphs. |
| | Multiple Choice | Asks students to choose from several answers (could be given as words, pictures, or equations) or answer "yes/no" questions (or questions that implicitly only have two answers). |
| | Ranking Task | Asks students to rank items (e.g. from most to least). |
| | Pie Chart | Asks students to draw a pie chart. |
| | Bar Chart | Asks students to draw a bar chart. |
| Evidence | Evidence | Asks students to use data to support a claim. Students should produce both claims and evidence. |
| Revisiting: Asks an initial question, then addresses the same question a second time. Does not include predictions. | Revisit with Reasoning | Any revisiting pattern that requires students' reasoning and doesn't fit the other strategies. A common wording for this strategy would be: "is your answer consistent with …". |
| | Procedural | Asks any revisiting question that uses traditional style procedures, such as plugging numbers into a formula. |
| | Check Printed Document | Asks students to check their answers against a printed document, such as a photograph or table provided by instructor. The word "consistency" may be used. |
| | Statement to Agree or Disagree | Statement to agree or disagree with (or in what way do you agree or disagree). Involves one or more statements often attributed to fictitious students for students to agree or disagree. |
| | Telling Answer | The activity or instructor tells students the answer to the question (or a set of questions) after the students answer the question. |
| | Revisit by Video File | Student checks their answer against a video file. |

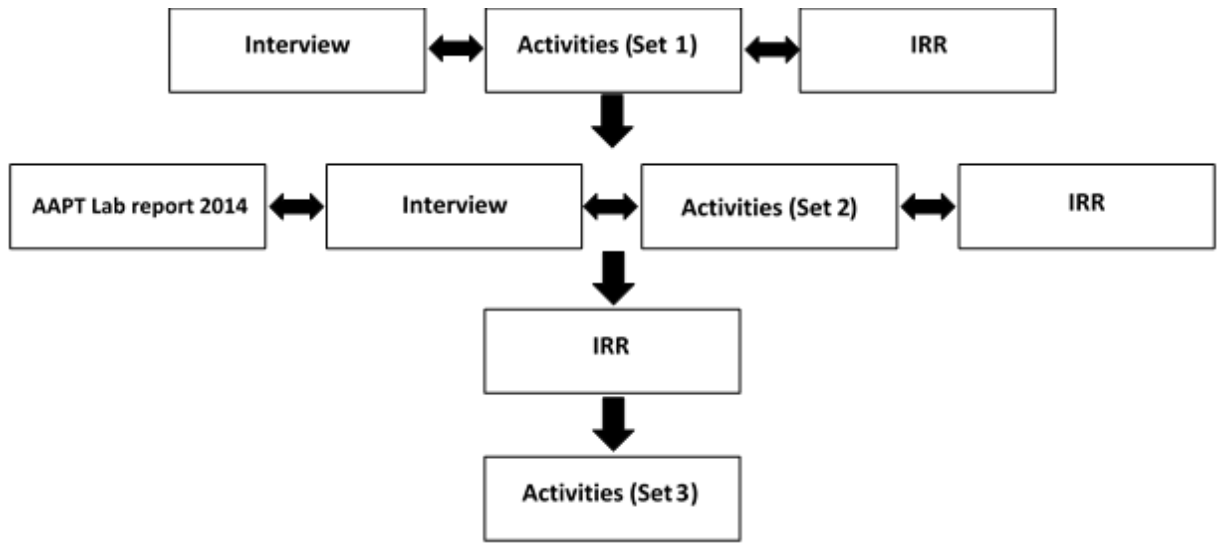

**Figure 1: Four steps of the coding process. Step 1: coding of interviews and 25 different activities and evaluating by IRR. Step 2: coding of AAPT lab report 2014 and 40 additional randomly-chosen activities, recoding of interviews, and evaluating by IRR. Step 3: randomly choosing one activity from each research-based curriculum and evaluating by IRR. Step 4: coding of five additional randomly chosen activities from each curriculum.**

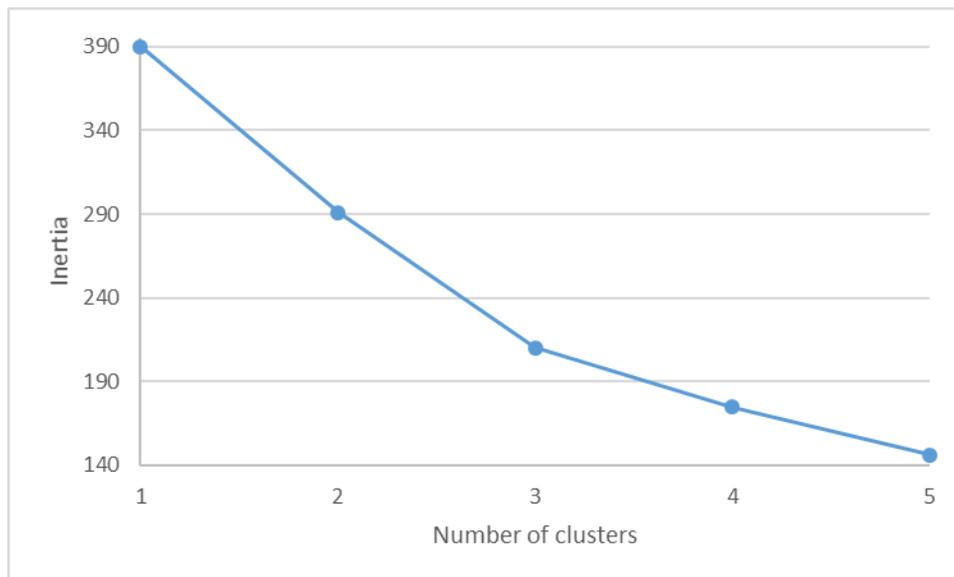

**Figure 2: Graph of smallest inertia (arbitrary units) achieved for each number of clusters. Smaller inertia corresponds to more compact clustering. As expected, the inertia decreases whenever the number of clusters is increased.**

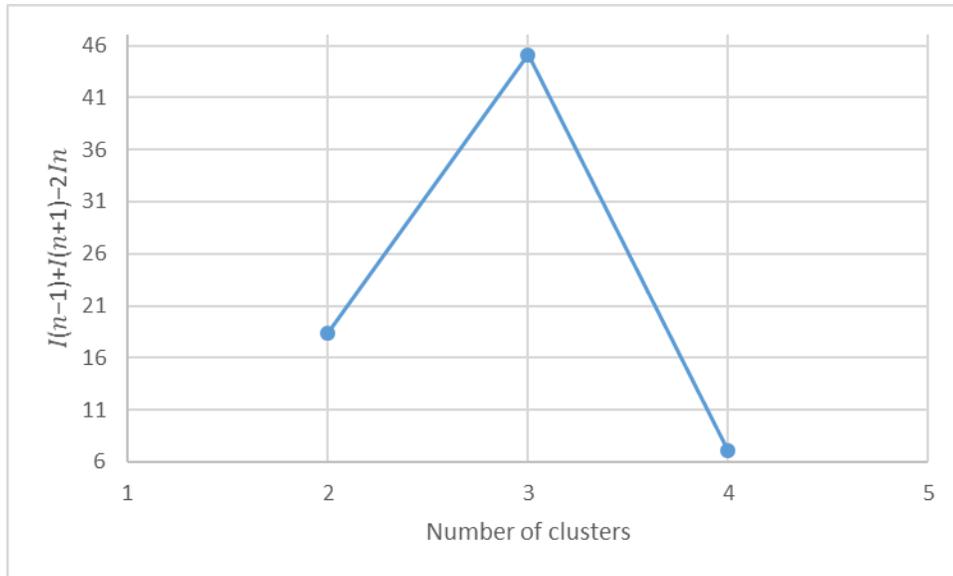

**Figure 3: Graph of decrease in inertia when increasing from N-1 to N clusters minus decrease in inertia when increasing from N to N+1 clusters (arbitrary units) vs. number of clusters. Elbow method analysis exhibits the highest value when increasing the cluster number has the largest relative impact which occurs for N=3 in our data.**

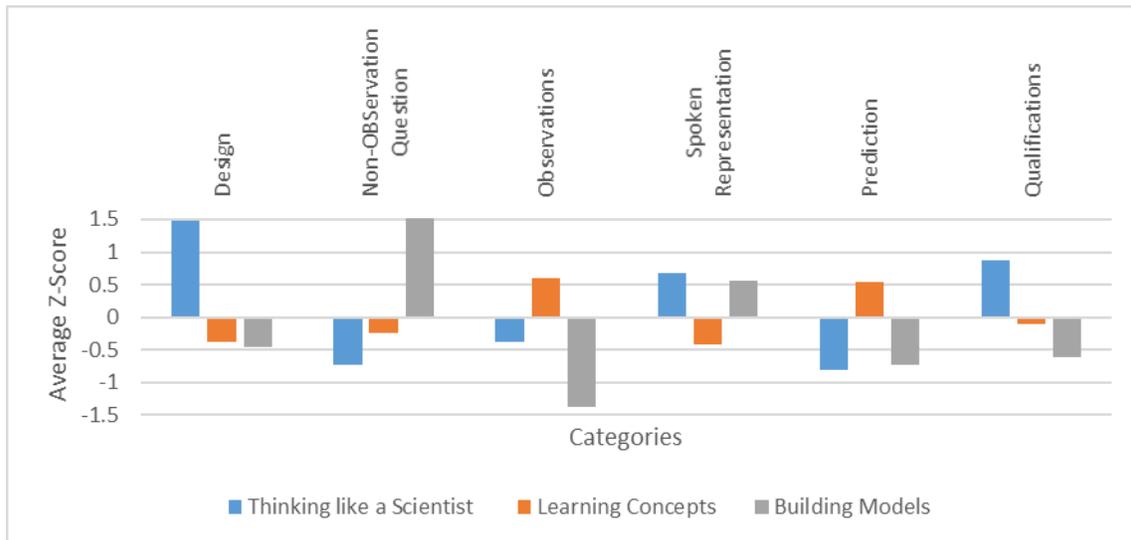

**Figure 4: Average Z-Score of each cluster for each of the analyzed features (Design, Non-observation Question, Observation, Spoken Representation, Prediction, and Qualifications).**